# Biphoton transmission through non-unitary objects

Matthew Reichert,[*] Hugo Defienne, Xiaohang Sun, and Jason W. Fleischer[†]

Department of Electrical Engineering, Princeton University, Princeton, NJ, 08544, USA

Email: [*]mr22@princeton.edu  [†]jasonf@princeton.edu

**Abstract**
Losses should be accounted for in a complete description of quantum imaging systems, and yet they are often treated as undesirable and largely neglected. In conventional quantum imaging, images are built up by coincidence detection of spatially entangled photon pairs (biphotons) transmitted through an object. However, as real objects are non-unitary (absorptive), part of the transmitted state contains only a single photon, which is overlooked in traditional coincidence measurements. The single photon part has a drastically different spatial distribution than the two-photon part. It contains information both about the object, and, remarkably, the spatial entanglement properties of the incident biphotons. We image the one- and two-photon parts of the transmitted state using an electron multiplying CCD array both as a traditional camera and as a massively parallel coincidence counting apparatus, and demonstrate agreement with theoretical predictions. This work may prove useful for photon number imaging and lead to techniques for entanglement characterization that do not require coincidence measurements.

## 1. Introduction

High-dimensional entanglement in the continuous variables of transverse position and momentum holds potential for many quantum processes and applications, such as quantum information [1-4], imaging [5,6] and lithography [7,8]. Typically, studies of entangled photon pairs, such as those generated via spontaneous parametric down-conversion (SPDC), involve post-selecting the two-photon portion from coincidence measurements to eliminate potentially large singles count rate arising from noise and losses. In quantum lithography, both photons are transmitted through an object, and then imaged onto a multiphoton-absorbing photoresist, where it may produce higher resolution features compared to classical coherent light and with greater contrast than classical incoherent light [7]. This benefit arises from the spatial correlation of the photons, where entangled pairs are localized together.

Within analyses of quantum lithography, and quantum imaging in general, loss in the imaging system is treated as an undesirable feature, and is often neglected. However, there is unavoidable loss from the object to be imaged which must be accounted for in a complete description of the system. In general, objects have a field transmission profile $|t(\boldsymbol{\rho})| \neq 1$, and are therefore non-unitary, meaning the form of the quantum state of light changes upon transmission. For pure biphoton state illumination, the transmitted state is composed of an attenuated two-photon term—the residual from the loss of either or both photons of a pair—and a single-photon term generated when only one photon of a pair is absorbed. Generation of single-photons from biphoton state is a well-known process used for heralded single-photon sources [9]. In imaging applications, the single-photon term is ignored, as only the detected two-photon portion of the state of interest is measured by coincidence counting. In the present work, we demonstrate that measuring the remaining single-photon term allows access to information about both *the object* and the *spatial entanglement properties* of the incident pairs.

Properties of biphoton states are conventionally characterized via coincidence counting using pairs of single photon detectors. When photons are generated in the same spectral and polarization mode, their features only depend on transverse spatial coordinates. A coincidence image can be constructed by scanning the detectors in the transverse plane [10-12]. Recently, the use of single-photon sensitive pixel-array detectors to measure coincidences—such as intensified

[13,14] and electron multiplying (EM) CCD cameras [15,16]—has substantially increased the potential of this type of source for imaging purposes. Such cameras have previously been used to characterize spatial entanglement of photon pairs [17-19] and to demonstrate the EPR paradox [15,16,19]. In the present work, we exploit these detection techniques with an EMCCD camera to study the transmission of spatially entangled photon pairs through non-unitary objects, i.e., those with loss ($|t(\boldsymbol{\rho})|^2 \neq 1$).

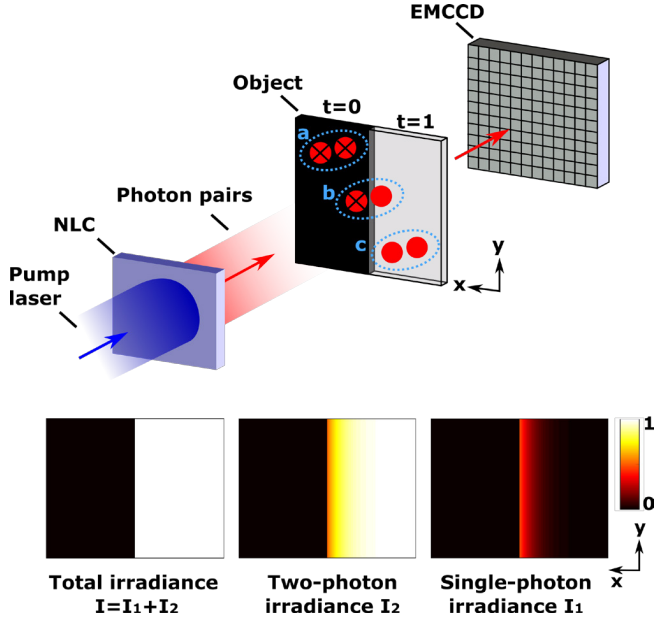

**Figure 1. Single-photons created by illumination of an absorptive object with entangled photon-pairs.** (Top) Biphotons generated by pumping a nonlinear crystal (NLC) exhibit spatial correlations. When they reach the object, three possibilities exist: (a) both photons absorbed, (c) both transmitted, or (b) only one of the two photons gets transmitted which occurs near to the edge. (Bottom) Irradiances after propagation of spatially correlated photon-pairs through the object. The total irradiance $I$ refers to all transmitted photons, regardless of their number state. The two-photon irradiance $I_2$ represents the spatial distribution of photons in a two-photon number state (biphotons). $I_2$ differs from $I$ due to absorption of one photon of a pair near the edge, which leaves transmitted photons in a single-photon state that does not contribute to $I_2$. The difference $I_1 = I - I_2$ is the distribution of photons in a single-photon state, that depends on both on the object properties and spatial correlation of the photon pairs.

The massively parallel capability of EMCCD's to detect the two-photon coincidence images has previously been used to measure correlation properties of the entangled photon pairs [15,16,19,20]. In the present work, this approach is extended to reconstruct the two-photon irradiance, $I_2(\boldsymbol{\rho})$, of the state, which is proportional to the marginal probability of detecting one photon of a pair at position $\boldsymbol{\rho} = x\hat{\mathbf{x}} + y\hat{\mathbf{y}}$. It corresponds to the image formed on the camera by only accumulating paired photons. For a pure biphoton source and unitary propagation, the two-photon irradiance is the same as the total irradiance $I(\boldsymbol{\rho})$, measured by accumulating all the photons on the camera, since they are all in a two-photon state. However, losses may remove one photon of a pair by absorption, leaving a certain portion of single photons to be detected by the camera. This is the case when, for example, spatially correlated photon-pairs illuminate an edge object (Figure 1). Far from the edge, photons from an entangled pair are either both (a) absorbed or (c) transmitted where $|t| = 0$ or 1, respectively. Additionally, only one photon from a pair may be absorbed while the other is transmitted, thus changing the number state of the transmitted light to a single-photon state. Because the entangled photons are localized to within a small correlation length of one another (blue dashed ellipses), the part of the transmitted state with a single-photon exists only near the edge of the object (b). These single-photons are not detected in coincidence by the EMCCD camera, resulting in a difference between $I_2(\boldsymbol{\rho})$ and $I(\boldsymbol{\rho})$, as shown in Figure 1. The single-photon irradiance, given by the difference $I_1(\boldsymbol{\rho}) = I(\boldsymbol{\rho}) - I_2(\boldsymbol{\rho})$, is the contribution of single-photons to the total irradiance, and is non-zero near the edge. The decay of $I_1(\boldsymbol{\rho})$ away from the edge is proportional to the correlation width of the incident biphotons. Therefore, the single-photon irradiance contains information about both the optical system (the object) and the entanglement properties of the photon pairs (the correlation width).

## 2. Experimental Methods

In the experiment, biphotons are generated via a near-collinear type-I SPDC in a BBO crystal pumped by a 405 nm cw laser diode, and near-degenerate PDC is selected using narrow band-pass filters (FWHM 10 nm). As shown in Figure 2(a), photon pairs illuminate an object through either a near-field (nf) or far-field (ff) imaging configuration for spatial correlation or anti-correlation, respectively. The object plane is then imaged onto a single-photon-sensitive EMCCD camera (Andor iXon Ultra 897) which is used as a massively parallel coincidence-counting apparatus to measure both the total and the two-photon irradiances. The EMCCD consists of a 512×512 array of 16×16 μm$^2$ pixels, and runs a temperature of –85 °C to effectively eliminate dark counts, and read out is performed at 17 MHz with a 0.3 μs vertical shift time. The camera is operated in photon-counting mode, where a binary threshold is applied to each pixel of ~2.8 standard deviations above the mean noise level (mainly due to clock-induced charge) [21,22]. A large number of frames (~$10^5$-$10^7$) are collected, each with a peak mean light level of ~ 0.15 photons per pixel per frame, chosen to minimize false detections [21] and maximize the signal-to-noise ratio [23].

Spatial entanglement properties of the source are first characterized, without an object in the experiment, employing the technique described in [15,16,19]. Briefly, for spatially correlated biphotons, the auto-correlation of each frame is calculated and summed together to give a conditional probability distribution of the separation of coincidence counts. A background, consisting of accidental counts between non-entangled pairs and noise, is estimated by the sum of cross-correlation of successive frames, and is subtracted. The same procedure is done for anti-correlated biphotons, but the convolution is calculated to measure the conditional probability distribution of mean positions of coincidences. Figure 2(b-e) show spatial correlation and anti-correlation of the photon pairs in the object plane using the near-field (nf) and far-field (ff) configuration, respectively. For both cases, the two-photon wave function can be approximated by a double-Gaussian function of the form [19,24]

$$\psi_i(\boldsymbol{\rho}, \boldsymbol{\rho}') = N \exp\left(-\frac{|\boldsymbol{\rho} - \boldsymbol{\rho}'|^2}{8\sigma_-^2} - \frac{|\boldsymbol{\rho} + \boldsymbol{\rho}'|^2}{8\sigma_+^2}\right), \quad (1)$$

where $N$ is a normalization factor and $\sigma_\pm$ are the standard deviations in the sum and difference coordinates of the chosen configuration. In the near-field configuration, photons are spatially correlated ($\sigma_-^{nf} \ll \sigma_+^{nf}$) with a measured standard deviation in difference coordinates of $\sigma_-^{nf} = 14.8 \pm 1.5$ μm. In the far-field case, photons are anti-correlated ($\sigma_-^{ff} \gg \sigma_+^{ff}$) with a standard deviation in sum coordinate of $\sigma_+^{ff} = 32.7 \pm 1.2$ μm. This gives an EPR product $\sigma_-\sigma_{k,+} = (5.0 \pm 0.6) \times 10^{-2}$ (where $\sigma_- = \sigma_-^{nf}$ and $\sigma_{k,+} = k\sigma_+^{ff}/f$) which is an order of magnitude less than than the Heisenberg bound of 1/2 [15,16]. The corresponding Schmidt number is 400 ± 90, indicating the high degree of spatial entanglement [25].

An optical slit is introduced in the object plane, where the spatial (anti-)correlation exists. For both configurations, the two-photon irradiance is reconstructed from a set of $N$ images $F_n[i]$ by selecting only the coincidences due to entangled photon pairs and subtracting accidental coincidences in a manner similar to [16]. The two-photon irradiance is calculated via

$$I_2[i] = \sum_j \left(\sum_n F_n[i]F_n[j] - \frac{1}{N-1}\sum_{n,m\neq n} F_n[i]F_m[j]\right), \quad (2)$$

For a given pixel $i$, the first term of this sum determines the number of coincidence between all other pixels within the same frame. Because there are many pixels above threshold (~15%), there is a large contribution from accidental coincidences between photons from different pairs, photons and noise events, or two noise events. Since entangled pairs always arrive within a single frame, we may estimate the distribution of accidentals by calculating coincidences between different frames, represented by the second term in the sum. Subtracting the uncorrelated accidental coincidences leaves only the entangled photon part of the total image. The single-photon portion of the transmitted state is finally calculated by subtracting the biphoton portion from the total irradiance $I_1(\boldsymbol{\rho}) = I(\boldsymbol{\rho}) - I_2(\boldsymbol{\rho})$. In practice, since the system has does not have unit quantum efficiency, we normalize $I_2(\boldsymbol{\rho})$ to the same value as $I(\boldsymbol{\rho})$ in the transparent region of the object (i.e., where $|t(\boldsymbol{\rho})| = 1$). This may be done ahead of time as a calibration step, where measurements are performed without an object and all photons arriving at the camera are in a biphoton state.

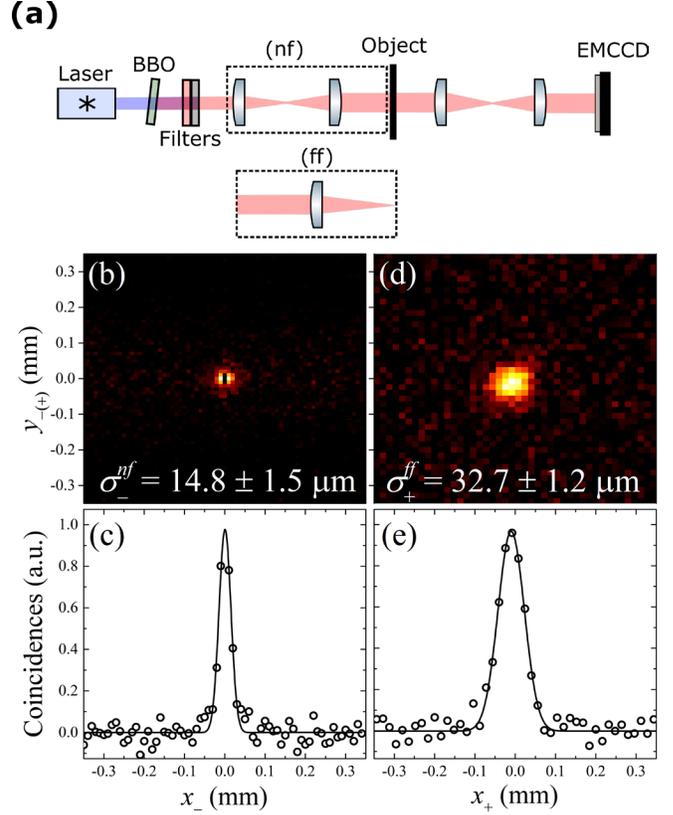

**Figure 2. Experimental apparatus used to generate entangled photon pairs.** (a) Biphotons are generated through a type I SPDC process in a 3 mm thick nonlinear crystal of beta barium borate (BBO) pumped by 405 nm cw laser (Laser). The pump is collimated with a radius of 840 μm (HW1/e²M). Long-pass and band-pass filters (Filters) are used at the output of the crystal to respectively block the pump photons and select only degenerate pairs centered at 810 nm (± 5 nm). The BBO crystal is slightly tilted to ensure near-collinear phase matching. An object is illuminated by the biphotons using either a near-field (nf) or far-field (ff) imaging configuration. (b-e) Biphotons exhibit spatial entanglement that can be characterized using the technique previously developed in [15,16]. Measurements of spatial (b, c) correlation and (d, e) anti-correlation calculated from auto-correlation and auto-convolution, respectively, of each measured image frame with background correlation subtracted. Here $x_\pm = (x \pm x')/\sqrt{2}$, and likewise for $y_\pm$, where $y_-$ applied to (b) and $y_+$ applies to (d). Black pixels in the middle of (b) are manually zeroed to remove charge smearing artifact [15], and are omitted from the fit in (c).

3. **Results**

Figure 3(a) shows the total irradiance measured in the near-field configuration (spatial correlation), with the individual two-photon and one-photon contributions in 3(b) and 3(c), respectively. $I_1(\boldsymbol{\rho})$ is only non-zero near the edges of the slit, where one photon from the pair is blocked. Figure 3(d) shows the vertical dependence (integrated over $x$) of each contribution to the total irradiance. The curves are predictions based on previously characterized incident biphoton wave function. The one- and two-photon irradiances are given by

$$I_1(\boldsymbol{\rho}) = |t(\boldsymbol{\rho})|^2 \int |\psi_i(\boldsymbol{\rho}, \boldsymbol{\rho}')|^2 [1 - |t(\boldsymbol{\rho}')|^2] d^2\boldsymbol{\rho}', \quad (3)$$

and

$$\begin{aligned} I_2(\boldsymbol{\rho}) &= \int |\psi_o(\boldsymbol{\rho}, \boldsymbol{\rho}')|^2 d^2\boldsymbol{\rho}' \\ &= |t(\boldsymbol{\rho})|^2 \int |\psi_i(\boldsymbol{\rho}, \boldsymbol{\rho}')|^2 |t(\boldsymbol{\rho}')|^2 d^2\boldsymbol{\rho}', \end{aligned} \quad (4)$$

respectively (see Appendix), where $\psi_o(\boldsymbol{\rho}, \boldsymbol{\rho}') = t(\boldsymbol{\rho})t(\boldsymbol{\rho}')\psi_i(\boldsymbol{\rho}, \boldsymbol{\rho}')$. In both cases, one photon from the pair is transmitted at $\boldsymbol{\rho}$, while the other at $\boldsymbol{\rho}'$ is either blocked or transmitted. In Eq. (3), the factor $[1 - |t(\boldsymbol{\rho}')|^2]$ in the integrand represents the losses that yields the single-photon portion. $I_2(\boldsymbol{\rho})$ falls off near the edge of the slit due to the decreased probability that both photons are transmitted as one is more likely to be blocked by the nearby edge. In the case where only one photon is transmitted, that pair no longer contributes to $I_2(\boldsymbol{\rho})$, but instead the surviving photon contributes to the one-photon irradiance $I_1(\boldsymbol{\rho})$ (Eq. (3)). Single photon states are created near both sides of the slit (blue) where one photon from the pair was blocked by the nearby edge. The $y$-dependence of $I_1(\boldsymbol{\rho})$ created near the edge can be approximated by (since $\sigma_-^{nf} \ll \sigma_+^{nf}$, and omitting the shift)

$$I_1(y) \propto \Theta(y) \left[1 - \text{erf}\left(\frac{y}{2\sigma_-^{nf}}\right)\right], \quad (5)$$

where $\Theta(x)$ is the unit step (Heaviside) function. Figure 3(d) shows agreement with the previously measured correlation width $\sigma_-^{nf}$. For a narrower correlation width the photon pairs would be localized closer together, and would be more likely to both be transmitted when they are closer to the slit edge. Measurement of $I_1(\boldsymbol{\rho})$, therefore, is a measure of the spatial correlation properties of the incident biphotons.

Spatially anti-correlated biphotons similarly result in a single-photon portion that depends on the anti-correlation width $\sigma_+$. However, they are not necessarily created equally near both edges. To demonstrate this we illuminate the slit with anti-correlated photon pairs (far-field configuration) and displace it slightly off center form the optic axis, about which the two entangled photons are centered. Figure 3(e-h) shows that single-photons are created only on the top edge of the slit near $y = 0.26$ mm, since their entangled pairs were localized near $y = -0.26$ mm and were blocked. At the other edge, photons that pass near $y = -0.20$ mm have pairs near $y = 0.20$ mm, within the transparent region of the slit. In direct measurements of $I_1(\boldsymbol{\rho})$, the asymmetry may be used to identify spatially anti-correlated biphotons since it is not present when they are correlated. Theoretical predictions via Eqs. (3) and (4) agree with the measured distributions and $\sigma_+^{ff}$ measured in Figure 2(d, e).

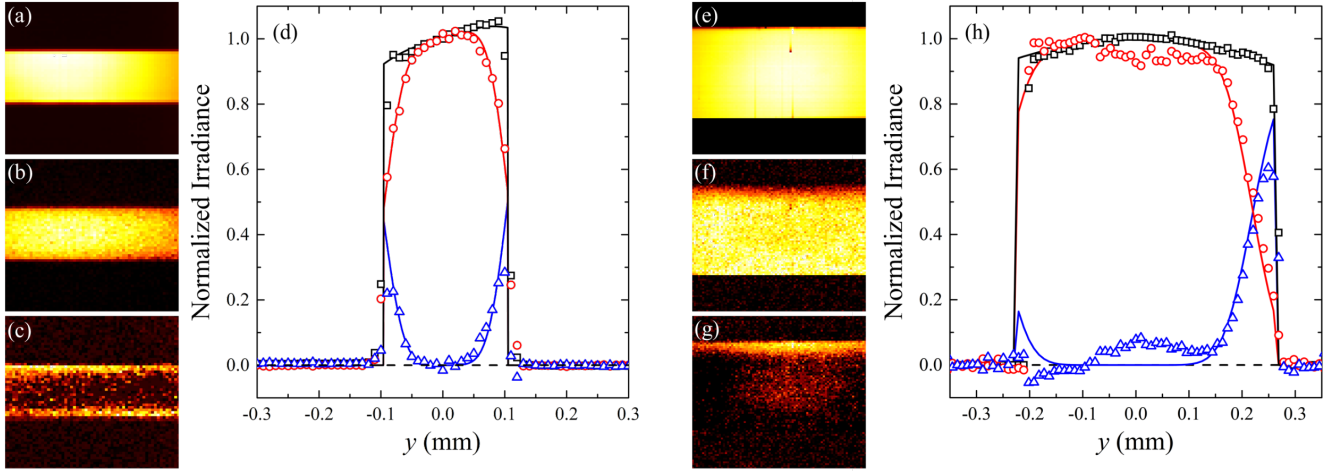

**Figure 3. Total, two-photon and single-photon irradiance of an optical slit illuminated by spatially correlated or anti-correlated photons pairs.** Spatial photon number measurements from biphoton illumination through a wide slit. (a-d) Images of slit illuminated with spatially correlated biphotons: (a) all photons, $I(\boldsymbol{\rho})$, (b) biphotons, $I_2(\boldsymbol{\rho})$, and (c) single-photons, $I_1(\boldsymbol{\rho}) = I(\boldsymbol{\rho}) - I_2(\boldsymbol{\rho})$. (d) Measured (shapes) $y$-dependence of irradiance in (a-c) with (curves) theoretical prediction. Slope of $I(y)$ (black) is due to the slit being positioned slightly off-center. (e-h) Spatially anti-correlated illumination with the slit positioned off center from the optic axis: (e) $I(\boldsymbol{\rho})$, (f) $I_2(\boldsymbol{\rho})$, and (g) $I_1(\boldsymbol{\rho})$. (h) Comparison of (shapes) measurement and (curves) calculation showing generation of single-photon state only on one side of the slit.

Our technique may also be applied to more complicated objects, since single-photon states are not only generated near step-like edges, but also anywhere $|t(\boldsymbol{\rho})| < 1$. For example, $t = 0.5$, still has a 50 % chance of generating a single-photon state, even with perfect spatial correlation where $\sigma_- \to 0$. In this case Eqs. (3) and (4) reduce to

$$I_1(\boldsymbol{\rho}) = |t(\boldsymbol{\rho})|^2(1 - |t(\boldsymbol{\rho})|^2)|\psi_i(\boldsymbol{\rho},\boldsymbol{\rho})|^2, \qquad (6)$$

$$I_2(\boldsymbol{\rho}) = |t(\boldsymbol{\rho})|^4|\psi_i(\boldsymbol{\rho},\boldsymbol{\rho})|^2. \qquad (7)$$

Note that in this case, since the photons are assumed perfectly correlated, the incident irradiance $I_i(\boldsymbol{\rho}) = |\psi_i(\boldsymbol{\rho},\boldsymbol{\rho})|^2$. Figure 4 shows measurements of a circular apodizing transmission mask which has unity transmission in the center that gradually falls to zero at the edges. The peak irradiance is displaced to where $|t(\boldsymbol{\rho})|^2 \approx 0.5$, where the one-photon portion of the state is expected to be maximized. From the incident irradiance and transmission function of the object (see Figure 4(a) and 4(b)), we can predict the single-photon portion of the transmitted state with Eq. (6), which is shown in Figure 4(c). The total, two-photon, and one-photon portions of the transmitted irradiance are shown in Figure 4(d), 4(e), and 4(f), respectively. The measured one-photon portion of the transmitted state agrees well with the theoretical prediction.

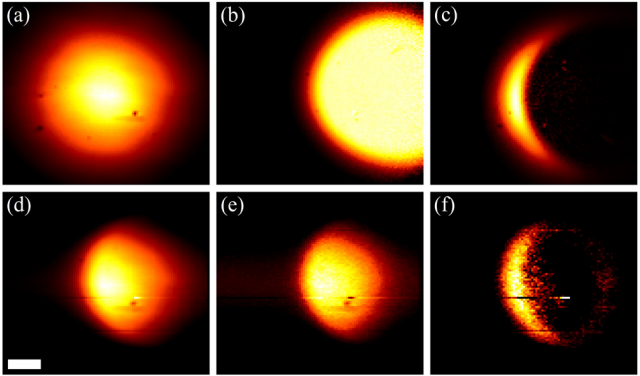

**Figure 4. One- and two-photon irradiances from an apodization mask illuminated by spatially correlated biphotons.** Total irradiance $I(\boldsymbol{\rho})$ measurement (a) without and (d) with apodizing mask, which are used to determine (b) the objects transmission function $|t(\boldsymbol{\rho})|^2$. (c) Predicted one-photon portion of the state based only on (a) and (b) via Eq. (6). (e) Measured $I_2(\boldsymbol{\rho})$ and (f) calculated $I_1(\boldsymbol{\rho})$ showing agreement with prediction in (c). Scale bar is 1 mm.

### 4. Conclusions

We are able to measure both the one- and two-photon irradiances of the state generated by transmission of a biphoton through a non-unitary object, that is an object with $|t(\boldsymbol{\rho})| \neq 1$. This essentially grants photon number resolution, albeit only between one and two photons and with the requirement that the input state is a pure biphoton state. The technique relies on the ability to measure the entire biphoton probability distribution, from which the two-photon portion of the irradiance can be determined. This is made possible by the massively parallel coincidence counting capability of an EMCCD camera, while traditional point-scanning techniques would be very time consuming and thus impractical.

The single-photon portion has an interesting dependence on the object. For objects with sharp edges, in both the near-field and far-field of the SPDC crystal, the one-photon term appears mostly at these edges, and therefore its measurement may act as a sort of edge detector. For objects with smooth gradients of $|t(\boldsymbol{\rho})|^2$, the one-photon portion becomes most prevalent where $|t(\boldsymbol{\rho})|^2 = 0.5$.

This technique may potentially be useful in applications that require number state discrimination, particularly with spatial resolution. For example, it would allow simultaneous measurement of both one- and two-photon absorption distributions of an object, or alternatively their spectra [26,27]. In this case, the two-photon portion of the transmitted state is attenuated by both one- and two-photon absorption, but single-photons are only generated by one-photon absorption.

Our approach may have interesting implications for quantum imaging and lithography, where entangled photons are anti-correlated in the far-field. An aperture stop in this plane would not only limit the resolution of such a system, but would also change the quantum state to generate some number of single-photons with high transverse momentum. In a quantum lithography system, loss would obviously reduce two-photon absorption rate, but the generation of single photons may also contaminate the image and reduce contrast. Such complications may warrant further exploration for practical implementations. Other potential applications

Remarkably, $I_1(\boldsymbol{\rho})$ contains information about both the object and correlation width of the entangled photon pair. In principle, this means that information about the spatial correlation of biphotons, and thus their degree of entanglement and EPR criterion, may be measured exclusively from single-photon measurements, i.e., without coincidence counting. Unfortunately, in order to obtain $I_1(\boldsymbol{\rho})$, $I_2(\boldsymbol{\rho})$ is first measured via coincidence counting techniques as presented here, which itself provides a measure of the correlation width. Direct measurement of $I_1(\boldsymbol{\rho})$, without coincidence counting and post-selection, may be possible with selective elimination of the two-photon term in the form of a strong nonlinear loss, such as up-conversion or two-photon absorption [28,29], to guarantee that no biphotons arrive at the camera. A recent technique using interferometric methods can accomplish this in the far-field [30]. However, since the method presented here operates equally well in both the near-field and far-field, such a system may allow demonstration of the EPR paradox without coincidence counting.

## Acknowledgments

This work was supported by the Air Force Office of Scientific Research.

## Appendix

Here we calculate the one- and two-photon portions of the transmitted quantum state of light arising from biphoton illumination in general, that is, derive Eqs. (3) and (4). The state generated via spontaneous parametric down-conversion (SPDC) in transverse spatial coordinates is (neglecting the vacuum term) [24]

$$|\Psi\rangle = \int \psi(\boldsymbol{\rho},\boldsymbol{\rho}')\hat{a}^\dagger(\boldsymbol{\rho})\hat{a}^\dagger(\boldsymbol{\rho}')d^2\boldsymbol{\rho}d^2\boldsymbol{\rho}'|0\rangle. \quad (8)$$

where $\hat{a}^\dagger(\boldsymbol{\rho})$ is the creation operator, and $\boldsymbol{\rho} = x\hat{\mathbf{x}} + y\hat{\mathbf{y}}$. The biphoton wave function $\psi(\boldsymbol{\rho},\boldsymbol{\rho}') = \langle 0|\hat{E}(\boldsymbol{\rho})\hat{E}(\boldsymbol{\rho}')|\Psi\rangle$, where $\hat{E}(\boldsymbol{\rho}) \propto \hat{a}(\boldsymbol{\rho})$ is the positive-frequency component of the electric field operator [31-33]. The biphoton wave function transmitted through an object is

$$\psi_o(\boldsymbol{\rho},\boldsymbol{\rho}') = t(\boldsymbol{\rho})t(\boldsymbol{\rho}')\psi_i(\boldsymbol{\rho},\boldsymbol{\rho}'). \quad (9)$$

To analyze the evolution of the quantum state as it is transmitted through a non-unitary object [34], we consider the object to be made up of a collection of beam splitters, each with two input ports and two output ports. Rather than the traditional cube beam splitter, we instead consider a thin object with the input ports as light incident from either direction, and the output ports the transmitted and reflected directions (see Figure 5). The quantum mechanical description of a beam splitter describes the coupling of annihilation operators $\hat{a}_j$ and $\hat{b}_j$ between the ports, where $j = \{i,o\}$. The properties of a beam splitter relate the creation and annihilation operators at the output ports of the beam splitter to those of the input ports [35,36]

$$\begin{aligned}\hat{a}_o &= t\hat{a}_i + r\hat{b}_i, \\ \hat{b}_o &= t'\hat{b}_i + r'\hat{a}_i.\end{aligned} \quad (10)$$

where $t$ and $r$ are the transmission and reflection amplitudes, with

$$|t|^2 + |r|^2 = 1, \quad (11)$$

and likewise for the primed varieties. In our case, we have an object with spatially varying transmission and reflection functions, $t(\boldsymbol{\rho})$ and $r(\boldsymbol{\rho})$.[1] To account for this, we also let the operators have spatial variation as well; $\hat{a}_j \to \hat{a}_j(\boldsymbol{\rho})$ and $\hat{b}_j \to \hat{b}_j(\boldsymbol{\rho})$.

---

[1] Here, we take reflection as a stand in for any loss. Since we neglect the reflected photons from the measurement (in fact neglect ports $\hat{b}$ entirely), we would get identical results if we explicitly included absorption as well. For more information see [35].

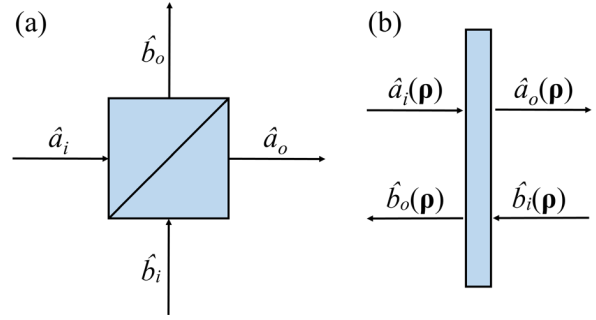

**Figure 5.** Quantum description of beam splitters relating the input and output creation operators for (a) traditional cube beam splitter, and (b) a semi-transparent object with spatially dependent transmission.

Let us now send the SPDC state into a beam splitter, where both photons enter the same input port, and determine the output state. The output state is then related to the input via Eqs. (10), which yields $\hat{a}_i(\boldsymbol{\rho}) = t(\boldsymbol{\rho})\hat{a}_o(\boldsymbol{\rho}) + r(\boldsymbol{\rho})\hat{b}_o(\boldsymbol{\rho})$. Taking the SPDC state (Eq. (8)), with both photons into the same input port of the beam splitter, $\hat{a}^\dagger(\boldsymbol{\rho}) \to \hat{a}_i^\dagger(\boldsymbol{\rho})$, the product of the two creation operators becomes

$$\begin{aligned}\hat{a}_i^\dagger(\boldsymbol{\rho})\hat{a}_i^\dagger(\boldsymbol{\rho}') &= t(\boldsymbol{\rho})t(\boldsymbol{\rho}')\hat{a}_o^\dagger(\boldsymbol{\rho})\hat{a}_o^\dagger(\boldsymbol{\rho}') \\ &+ t(\boldsymbol{\rho})r(\boldsymbol{\rho}')\hat{a}_o^\dagger(\boldsymbol{\rho})\hat{b}_o^\dagger(\boldsymbol{\rho}') \\ &+ r(\boldsymbol{\rho})t(\boldsymbol{\rho}')\hat{b}_o^\dagger(\boldsymbol{\rho})\hat{a}_o^\dagger(\boldsymbol{\rho}') \\ &+ r(\boldsymbol{\rho})r(\boldsymbol{\rho}')\hat{b}_o^\dagger(\boldsymbol{\rho})\hat{b}_o^\dagger(\boldsymbol{\rho}').\end{aligned} \quad (12)$$

The output state is the sum of four terms. The first term represents the part of the probability amplitude where both photons are transmitted. The second and third terms each have one photon transmitted and one reflected, and are identical since the state is symmetric under exchange $\psi(\boldsymbol{\rho}_1,\boldsymbol{\rho}_2) = \psi(\boldsymbol{\rho}_2,\boldsymbol{\rho}_1)$. The forth term has both photons reflected.

We are interested in examining the photon number of output port $\hat{a}_o$ (transmitted) independently of that of $\hat{b}_o$ (reflected). The first term, where both photons go to $\hat{a}_o$ remains unchanged, and leads to the two-photon term. The irradiance of the two-photon term is given by the marginal of the output biphoton wave function (Eq. (4)). The second and third terms in Eq. (12), however, each only have one photon in $\hat{a}_o$. For the spatially multimode incident biphoton wave function, the single photons transmitted are incoherent, and described by an irradiance distribution given by Eq. (2), with $|r(\boldsymbol{\rho})|^2 = 1 - |t(\boldsymbol{\rho})|^2$.